\documentclass[preprint,12pt]{aastex}

\slugcomment{In JAAVSO 30, 23 (2001)}

\shorttitle{Hot Flashes on Miras}
\shortauthors{Willson \& Struck}

\begin{document}


\title{Hot Flashes on Miras}


\author{L. A. Willson\altaffilmark{1}, and Curt Struck\altaffilmark{2}}  
\affil{Department of Physics and Astronomy Department, Iowa State University,
Ames, IA 50011}


\altaffiltext{1}{Dept. of Physics and Astronomy,
  12 Physics Bldg., Iowa State Univ., Ames, IA 50011,
lwillson@iastate.edu} 
\altaffiltext{2}{curt@iastate.edu}


\begin{abstract}
Short time scale variations have been reported for a few Miras,
including sudden 0.2 magnitude or more brightening in the visual
lasting from a few hours to a few days.  Are these flashes real?  Are
they hot?  What are they?  Almost all the natural time scales for
variation in the atmosphere and wind of a Mira variable are months to
years long.  What could cause such short-term variations?  One
intriguing possibility is that the variations are associated with the
interaction of a Jovian planet with the time-dependent outflow of a
Mira wind.  Here we discuss observable features of such an interaction
in terms of order-of-magnitude estimates and phenomenology, and make
one clear prediction requiring observational followup.  Future work
needs to include both theoretical calculations and the development of
systematic methods for searching for such events.
\end{abstract}




\section{Introduction: The Normal Variability of Miras}

The light curves of Miras are characterized by large amplitudes (2.5
visual magnitudes or more), long periods (typically a year and ranging
from about 100 to over 600 days), and relatively stable periodicity,
but with erratic cycle-to-cycle variations and small period changes on
decade-long time scales.  The typical light curve has a more rapid
rise than its decline and shows nearly straight-line variation in
magnitude with time on the declining branch (corresponding to an
exponential decay in brightness).  There are often extended ``bumps''
occurring on the rising or descending branches at about the same phase
every cycle. The timing and magnitudes of the bumps and of maximum
light are typically variable on a scale of $10-20 \%$, both phenomena
most likely associated with the emergence of shocks through the
visible-light photosphere (see e.g., \citet{maf95}) and quite
unrelated to the topic of this paper.

These normal variations can be understood, at least qualitatively,
based on models such as those by G. H. Bowen (\citet{bow88}; reviewed
by \citet{wil00}).  The star pulsates radially in a manner similar to
that of the Cepheids and RR Lyrae stars, except that in this case it
is a combination of hydrogen and helium (first) ionization that
provides the ``valve'' for the pulsation (\citet{woo74},
\citet{kee70}).  In the atmosphere, these waves build up to become
shock waves that propagate out through the atmosphere and into the
wind. There are typically two shocks formed per cycle, providing a
natural explanation for the ``bumps'' and perhaps also for those
objects showing double maxima.  An outflow is generated, either by the
pulsation alone or, in some stars, also by radiation pressure on dust
grains formed in ``refrigerated'' zones in the atmosphere.

Mira variables have pulsation periods around one year, and this is
also about how long it takes material and shocks to travel, at
typical speeds of $5-10 \ km/s$, one to two AU, or about one stellar
radius.  This is then the natural time scale for variation.  

Other mechanisms that have been suggested as possibly influencing the
visual brightness include large-scale convective cells and solar-type
flares.  The natural time scale for variations linked to convection
will also be of the same order as that for the pulsation. Solar-type
flares, although quick, would need to be very bright - orders of
magnitude brighter than flares observed on any other single star - to
contrast with these very luminous stars. There is no evidence for the
kinds of magnetic fields that would be required. (For a discussion of
this point, see \citet{sok02}.)

Not seeing a plausible mechanism for short-time-scale high-contrast
events, we have been very slow to accept reports of these, at least
until some systematic effects could be established in a homogeneous
data set. Such data are now emerging, and with further reflection we
are also seeing some interesting possibilities for what the data may
be telling us.

\section{Mira ``Flash'' Events From the Literature}

A few years back, Bradley Schaefer collected some data suggesting that
there were shorter period variations on some Miras \citep{sch91}.  His
collection of events was rather heterogeneous, with a factor of 100
variation in time scale, changes in magnitude from 0.4 to 1.4, phases
from maximum to mimimum, appearance on both rising and falling
branches, and methods including visual, photographic, and radio
observations. Three stars are listed for repeat events in his list,
but unlike the de Laverny report discussed below, none of them showed
a pattern of repeating at the same phase. So we did not take this
report very seriously. A more recent paper by \citet{maf95},
apparently discussing the same thing, turns out on closer inspection
to be focused on the behavior near maximum light.  However,
\citet{del98} found possibly short-term variations in the Hipparcos
data for $15\%$ of the Miras observed.  That is a large percentage,
although note that some of their cases were decreases in brightness
and not ``flashes''.  Again here, the data are not compelling by
themselves; most instances are single-datapoint departures from the
expected light curve.

There were some patterns noticed by de Laverny et al. that we find
very suggestive.  First, these events were systematically found away
from maximum, most likely again because the star is dimmer rather than
indicating that the phenomenon is phase-dependent.  This is a natural
result if the flash has a limited visual brightness and only shows up
when the contrast is highest (when the star is faintest).  Second, the
stars showing flashes were systematically among the latest in spectral
type, perhaps because the later types usually have bigger amplitudes
(and are thus fainter near minimum light) or because the later types
typically also have more massive winds.  Finally, there were several
stars that showed repeated flashes, with these showing up twice at the
same phase in each star (but at different phases for different
stars). De Laverny et al. said ``...In nearly half of these cases, the
variations occur {\it {at virtually the same phase:}} RT Boo, X Hya, W
Vel. In two instances (X Ant and RX Mon) the phases are almost
symmetric with respect to minimum of brigthness. The two variations of
Z Oct belongs to one of these two cases. Furthermore, two of the three
variations of X CrB and AM Cyg have the same phase, whereas the third
one of X CrB is symmetric with respect to $\phi = 0.5$.  However, the
number of stars with two detections is too small to draw conclusions
about a link between rapid variations and phases. One can only say
that, {\it {if such a link exists, the phase of occurance depends on
the stars.''}}

One way to check the reality of these events is to see what the AAVSO
data look like.  However, we run into a problem here.  The probable
duration of the events, deduced from the spacingbetween observations
in nearly all cases because just one data point was high, is a few
hours to several days.  Phenomena on such a short time scale will not
show up at all obviously in AAVSO data, since each observer observes
a given star at most once per week (for good reasons) and the
observations typically have a moderately broad scatter anyway at such
red colors.  We have looked at the AAVSO light curves for the times of
flashes reported by de Laverny et al. and found no compelling evidence
for or against these events in the data. 

Two stars were found by de Laverny et al. to have 3 flashes.  The
flash data for these two stars is reproduced in Table 1. Note that the
phase of repeat is not minimum light, but 0.1 to 0.2 in phase earlier. 

\begin{deluxetable}{lccc}
\scriptsize
\tablecaption{Multiple events for two stars as reported 
by de Laverny \it{et al.} 1998.
\label{tbl-1}}
\tablewidth{0pt}
\tablehead{
\colhead{JD}&
\colhead{Phase} &
\colhead{$|{\Delta}mag|$} &
\colhead{Duration}\\
2440000+ &&& in hours
}
\startdata
\cutinhead{$X\ CrB,\ P=241^d.2, Sp=M5.0-M7.0,$ phase of min = 0.54}
8116.0& 0.41& 0.23& 2.48\\
8607.3& 0.42& 0.40& 6.06\\
8639.8& 0.55& 0.58& 2.13\\
\cutinhead{$AM\ Cyg,\ P=370^d.6, Sp=M6.0-?,$ phase of min = 0.6}
\\
8302.5& 0.40& 0.56& 138.67 \\
8446.7& 0.78& 0.32& 2.48\\
9036.6& 0.38& 1.11& 115.27
\enddata

\end{deluxetable}

A couple of others were found to have two events, again at nearly the
same phase.  This repetition at a fixed phase suggests that whatever
is creating the event is located at some fixed distance from the
center of the star, so that the pulsation-induced shock comes through
at about the same phase in every cycle.  One may also speculate that
the preference for phases near 0.4 in these two stars could result
from the competing demands that the stellar flux be low (sooner after
maximum light) and that the shock still be in a region with relatively
high density (closer to minimum light).  With greater sensitivity we
would expect to see events at earlier and earlier phases.



\section{Energetics and Contrast}

In their paper, de Laverny et al. show a histogram of the phases of
the detection of short-term events in the Hipparcos data.  It
resembles very closely the light-curve of a Mira variable plotted
``upside down'', with brightest points at the bottom.  This supports
the idea that the preference for some phases is mainly that it is
easier to see a flash when the star is faint.  However, we note that
some of their events were dimmings, and this argument does not explain
why those should be concentrated towards minimum. They did not
indicate which events were dips and which were peaks; we are looking
back through the data to check this.

Most of the energy radiated by a Mira comes out in wavelengths between
1 and 5 microns.  Near visual maximum, the visual brightness is much
greater than the Sun's, but near minimum, particularly for a
large-amplitude star, the visual brightness may be only about the same
as the Sun's, even though the star is still radiating several times as
much power as the Sun does.  If we are to see an event with a contrast
of 0.1 or 0.2 magnitudes, then this event must by itself be emitting
ten to twenty percent of the visual flux.  If the peak of the flash
output is in the visual (a ``hot'' flash of 5000-10,000K equivalent
temperature) the we get the minimum energy that must go into the flash
for us to see it.

Figure 1 shows the mean light curve of $\chi$ Cyg in units of log
($F_{visual}/F_{visual, Sun}$).  To get this, we took a distance to
$\chi$ Cyg of about 100 pc as is indicated by its period and the P-L
relation for Miras (see discussion in \citet{wil00}).  We transformed
this to power units using the magnitude formula

log(Fvis/Fvis, Sun) =3D -0.4(Mvis-Mvis, Sun)

\begin{displaymath}
log(F_{vis}/F_{vis,Sun}) = -0.4 (M_{vis} - M_{vis,Sun}).
\end{displaymath}

The other three curves in Figure 1 are the result of adding 0.01, 0.1
and 1.0 times the brightness of the Sun to the flux of the Mira.  It
is clear that it is much easier to see an event that occurs when the
star is faint; when the star is bright the contrast is too small,
giving a small change in the magnitude. From Figure 1 we see that a
flash with visual output equal to our Sun's would be visible over a
large range of phases for $\chi$ Cyg - roughly 0.2 to 0.8 - while one
with a visual output of 10\% that of our Sun would be visible from
0.35 to 0.7. The minimum energy required to get 10\% of the Sun's
visible output is about $0.1 L_{Sun}$ for a color temperature of about
6000K. 

The observational data may then be summarized as follows: 

* The events, if real, last from a few hours to several days.  

*The total power involved may be as low as 0.1 solar luminosity if we
take the ideal  ``hot flash'' condition (T = 6000K) but is likely to be more.

\section{Possible, Probable, and Improbable Models}

Any model that is to account for the ``hot flashes'' must give (a)
enough energy being released in a single flash, (b) flashes appearing
at about the same phase in more than one cycle, (c) flashes appearing
more often in the light curves of stars of later spectral types, and
(d) flashes lasting hours to a few days.

We know that there are short-term variations on the Sun, solar flares,
and that these show up with greater contrast in late-type main
sequence stars.  Could the ``flash'' events on Miras be solar-type
flares?  Solar flares result from the transfer of energy from the
magnetic field into particle motion and radiation. On the Sun, these
last from a few minutes to an hour or so and have power output
typically $\le$ or $\ll 0.01 \times L_{Sun}$. What determines the
power, color, and time-scale of a solar-type flare?  On the Sun,
flares are relatively unpredictable events, occurring where the
magnetic field crosses itself in a particular geometry.  The ultimate
cause of the flare is the windup of magnetic field lines in the Sun as
a result of differential rotation.  In a Mira, a windup of field could
occur as the result of convective motions, and might even yield the
necessary brightness, but this model doesn't naturally produce events
repeating at a single phase (different from minimum), and does not
lend itself to modeling at the present time because too many factors
are unknown. (See \citet{sok02} for a discussion of these and related
points.)

Another phenomenon that could give rise to a short-lived ``flash'' is
what we have termed ``shock coalescence''.  In some computer models of
Miras there is a tendency for some shocks to run slow and some to run
fast, so that two shocks sometimes coalesce.  This occurs as a result
of nonlinear instabilities in the atmosphere, and since there is no
communication from one part of the outflowing wind to another part
some distance around the star, this instability should give rise to
hot patches in the circumstellar outflow.  However, like the flares,
these events should not be periodic, and there are reasons to think
that they may be rare in real stars because as the models have become
more realistic the shocks have become more periodic.  Finally, the
``coalescence'' events in models occur some considerable distance from
the star, while to produce enough energy they would have to occur
quite close to the star.

Two other ideas come from looking at the chemistry of the atmosphere.
\citet{ste00} and \citet{ste01} suggested that there might be abrupt
changes due to ``molecular catastrophes'' that result from the fact
that CO is very good at radiating away energy.  When the gas is too
hot to have CO in it, then it also cools slowly.  As soon as CO forms,
the gas begins to cool quickly.  This could abruptly change the
opacity of the atmosphere, allowing more or less visible light to leak
out (like making smoke signals with a blanket).  Similarly, when dust
grains form, they pick up momentum from the stellar radiation and
begin to move away from the star; the gas is dragged along with the
dust as dust grains and atoms collide.  This expansion of the
atmosphere produces cooling which further increases the rate of dust
formation.  Both of these mechanisms are very interesting and I would
suspect are present in the real stars.  They also could show phase
stability, if they occur at a certain distance from the star where the
density is ``just right''.  However, neither one seems likely to
produced sudden and short-lived brightenings in the star; even if the
onset is quick, it should take a time more like a pulsation cycle to
get out of the ``hole'' of more CO or more dust. (Note, however, that
this kind of opacity effect may be of interest in explaining the
variations seen around maximum light.)

Our preference is to interpret the flash as being the result of some
phenomenon associated with a shock passing an orbiting Jovian planet
or brown dwarf companion star. One way to accommodate the short time
scale is to consider the time it takes something moving around 5 to 10
km/s (such as an outward moving shock) to cross the sphere of
influence of a planet.  The extreme times are 6 hours to 3 days; 6
hours at 10 km/s covers 216,000 km and 5 km/s for 3 days covers
1,296,000 km. For comparison, 216,000 km = just over $10^{-3}$ AU or
about 1/3 of the current solar radius. These match well with estimates
for the sphere of influence for Jupiter, taken as the sphere within
which the local escape velocity from Jupiter is greater than the wind
flow velocity. 

There are two ways that energy could be stored over a cycle and then
released abruptly in this context: (a) there could be some sort of
magnetic windup associated with the planet's interaction with the
wind; or (b) the energy could be stored as kinetic energy in a
circumplanetary disk. Aspects of the first model are discussed by
\citet{cun00}; we are concentrating on exploring the second variety. 

Consider the possibility that the ``event'' is caused by the
interaction of the wind with a planet.  The energy constraint is then
a strong constraint. Suppose two amounts of mass with total mass M are
colliding with a velocities of $v$ relative to their center of mass;
then the energy released will be about $\frac{1}{2} M{v^2}$ and this
must be at least as large as $\int L dt \ge (10^{32} ergs/sec) \times
(10^5 sec) = 10^{37} ergs$ if the event persists over one to three
days.  If $v$ is 5-10 km/s (typical speeds) then the minimum mass
involved is at least $10^{-8} M_{Sun}$. This value is about what you
would get if you could use the integrated mass loss of a typical Mira
over three days in all directions, $\int M_{Mira}dt \simeq (0.01 year)
\times 10^{-6}M_{Sun}/year) = 10^{-8}M_{Sun}$.  But we have already
argued that we have to be dealing with a small volume, implying that
only a small fraction of the wind is allowed to be involved.

We seem to have reached an impasse, but that would be a premature
conclusion. While taking just the amount of material available during
a few days of wind flow near a Jupiter would not yield enough energy,
the gravity of such a planet is such that it will not tend to create a
distant bow shock but rather will pull wind material down into an
accretion disk.  The direction of the circulation in the accretion
disk will depend on the direction of the planet's motion and on
whether the wind is flowing out from or falling back towards the star.
In any location at all close to a Mira, for about half the cycle
material is flowing out and for the other half, falling in.  So the
direction of flow reverses twice during a pulsation cycle, first
abruptly, when the shock passes, and later more gently, about half-way
between shocks.  When the direction of preferred orbital motion about
the planet changes, the new material will run head-on into the ``old''
material accreted during the other half-cycle.  The result could be a
rather sudden release of a lot of energy.  How much energy?  Simple
estimates show that this could yield about a tenth of a solar
luminosity for a moderately large Jovian planet to perhaps a solar
luminosity for a brown dwarf - about what we need.  We would expect to
see this for several cycles at about the same phase (when the
pulsation-shock reaches the orbit of the planet), and then perhaps not
see it for several cycles while the planet is behind the star.  We
would expect cycle-to-cycle variations in the brightness of the flash
as we view the interaction from different angles.  The time scale for
the ``event' could plausibly be about the time it takes all the
material in the accretion disk to make one orbit around the planet -
and this comes out to hours to a few days for reasonable parameters
for the planet. 

The above is the simplest model - one in which the energy is stored as
kinetic energy. It is also possible to store and release magnetic
energy in this situation, taking advantage of the winding up of the
magnetic field in the circumplanetary accretion disk. We are
concentrating for now on the simpler case, for our modeling, but note
that the fundamental predictions of this paper are the same whether
the release has a magnetic component or not. 

How can we test this hypothesis that the flashes result from the
interaction of the outflow with a Jovian planet? We are proposing to
carry out more detailed calculations in order to make some very
quantitative predictions about the character of the events and their
influence on the winds.  Meanwhile, there is one clear prediction that
our model suggests: As the planet orbits around the star, taking
perhaps ten years to do so, it will sometimes be between us and the
star and sometimes behind the star when the shock passes. We would
expect, then, to see a longer cycle (the orbital period of the planet)
in the appearance and disappearance of the flashes in at least some of
these stars. Over the orbital period, the flashes should appear,
strengthen, then weaken and disappear as the planet circles the
star. The flashes should appear at the same phase when they appear,
but would not be present in every cycle. To check this prediction, we
need first to identify a group of reliably flashing stars, and then
monitor them for 10 to 20 years in a manner calculated to detect and
follow flash events. This would be a new kind of observing program for
the AAVSO and we are working to set it up.

To return to the questions we asked at the beginning, concerning the
hot flashes:  Are they real?  We aren't sure, but there are some
interesting hints that they may be real.  Are they hot?  Short-lived
events that are not ``hot'' ($6000-10,000K$) are unlikely to show up in
contrast with the stellar brightness, so if they are real then they
are almost certainly hot.  What are they?  We have proposed an
interesting possibility, namely that these events could signal the
presence of a Jovian planet in orbit around a dying star.

\acknowledgments

\clearpage



\begin{figure}
\plotone{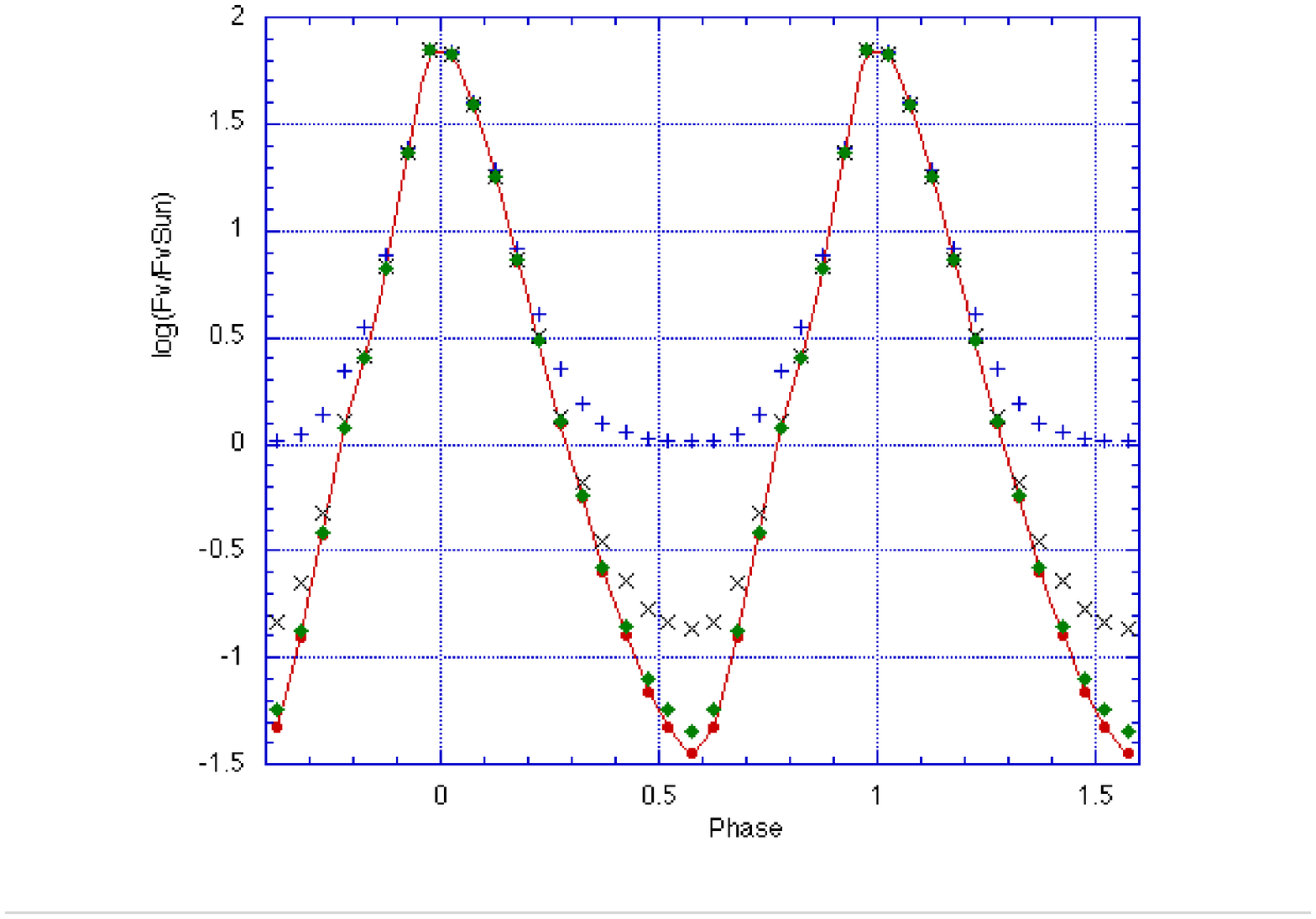}
\caption{Chi Cyg's visual light curve transformed to flux relative to the
visual flux of the Sun (solid red line) and with 0.01, 0.1, and 1
times the solar flux added.  An event with 0.01 times the visual flux
of the Sun (or 0.01 times LSun if the flux peak is in the visible) is
barely detectable near minimum with good photometry.  An event with
0.1times the flux of the Sun should b detectable from phase 0.35 to
0.7, roughly, but most readily near minimum.  An event with the same
visible flux as the Sun would be detectable over a slightly larger
range of phases - about 0.2 to 0.8 - and should ``knock your eye out''
near minimum.\label{fig1}}
\end{figure}





\end{document}